**Heuristic evaluations of back support, shoulder support, hand grip strength support, and sit-stand support exoskeletons using universal design principles.**

Alejandra Martinez, Laura Tovar, Carla Irigoyen Amparan, Karen Gonzalez, Prajina Edayath, Priyadarshini Pennathur, and Arunkumar Pennathur*

Physical, Information and Cognitive Human Factors Engineering Research Laboratory
Industrial, Manufacturing and Systems Engineering Department
University of Texas at El Paso, El Paso, Texas 79968-0521, USA

*Corresponding author; Email for corresponding author: apennathur@utep.edu

*Conflict of Interest*
The authors declare no conflicts of interest.

*Acknowledgments:*
The authors thank Dr. Panfeng Liang, a research statistician at the Data Analytics Lab, and Dr. Abhijit Mandal, Director of the Data Analytics Lab at the Border Biomedical Research Center, for their assistance in conducting the statistical analyses reported in this paper.

*Funding:*
This research was supported by a STAR grant from the State of Texas and funds from the College of Engineering at the University of Texas, El Paso.  The contents of this paper are solely the responsibility of the authors and do not necessarily represent the official views of the State of Texas or the University of Texas at El Paso.

**Heuristic evaluation of back support, shoulder support, hand grip strength support, and sit-stand support exoskeletons using universal design principles.**

**Occupational Applications (≤150 words)**

Our evaluation of four occupational exoskeletons using universal design principles revealed opportunities for design improvement. Our results indicate that designing exoskeletons for equitable use by all types of workers, including workers with disabilities, older workers, and women, remains a challenge. Assembling exoskeletons for use, donning these wearable devices, and doffing them pose challenges, particularly because of the strength, dexterity, reach, and balance requirements of the user. Workers with diverse capabilities may not be able to assemble or don these devices without additional support from another person. Exoskeleton designs can be improved to provide feedback on user actions, error prevention, and error recovery. For industry adoption, factors such as assembly and storage space demands, training needs, additional personnel to assist users, and personalization costs could pose significant barriers.

**Technical Abstract (≤300 words)**

Background (or Rationale): Occupational exoskeletons promise to reduce the incidence of musculoskeletal injuries; however, we do not know if their designs allow universal use by *all* workers. We also do not know how easy the tasks of assembling, donning, doffing, and disassembling exoskeletons are.

Purpose: The purpose of our study was to heuristically evaluate a back support, a shoulder support, a handgrip strength support, and a sit-stand exoskeleton for how well they are designed for universal use when assembling, donning, doffing, and disassembling the exoskeleton.

Methods: Seven evaluators used universal design principles and associated criteria to independently evaluate and rate four exoskeletons when assembling, donning, doffing, and disassembling the devices.  The rating scale was a Likert-type scale, where a rating of 1 represented not at all, and a rating of 5 represented an excellent design with respect to the universal design criteria for the task.

Results: The results indicate that providing perceptible information to the user, making the design equitable to use for a diverse set of users, making the design simple and intuitive to use with adequate feedback, and designing to prevent user errors, and when errors are made, allowing the user to recover quickly from the errors, were rated poorly. Assembling and donning tasks presented the most challenges.

Conclusions: For the industry to widely implement exoskeletons and exoskeletons to acquire the status of personal protective equipment (PPE), exoskeleton manufacturers must consider a wider range of users, address critical safety concerns when assembling and donning these devices, simplify designs to be a one-person operation, and consider industry barriers such as training needs and customization of devices in their design process. The exoskeleton-human factors research community should include diverse users in their evaluations and conduct usability, accessibility, and safety evaluations of these devices to provide design feedback.

1. **Introduction and background**

The incidence and prevalence of non-fatal occupational musculoskeletal disorders in the United States have recently increased in many industrial sectors. According to the most recent survey of employer-reported non-fatal workplace injuries in the United States (Bureau of Labor Statistics, 2022), the total number of reported injuries has increased by 6.3% from 2.1 million cases in 2020 to 2.2 million in 2021. Occupational exoskeletons have emerged as a promising solution to alleviate work-related musculoskeletal disorders in industrial workplaces and maintain worker productivity and safety while performing industrial tasks (Elprama et al., 2020; Howard et al., 2019, 2020; Kermavnar et al., 2021; Kim et al., 2018; Kuber et al., 2022, 2023; McFarland & Fischer, 2019; Medrano et al., 2023; Nussbaum et al., 2018; Papp et al., 2020; Reid et al., 2017). Originally developed for military applications and in rehabilitation settings, exoskeletons are "wearable devices that augment, enable, assist, and enhance physical activity through mechanical interaction with the body" '(Lowe et al., 2019). Market analysts predict that the exoskeleton market will reach $1.8 billion in value by 2025, indicating significant industry interest and support for adopting and implementing exoskeletons in the work environment (*Financial Results for Publicly Held Exoskeleton Companies, 2016 – 2022*, 2023). In the European Union, some exoskeletons have already earned the European Union conformity (CE) personal protective equipment (PPE) mark (*Laevo FLEX 3.0 is the first-ever exoskeleton Issued Personal Protective Equipment PPE CE Mark*, 2022), and it is likely that the US will soon follow suit.

Occupational exoskeletons hold remarkable promise for becoming an important ally to industry and workers for alleviating workplace injuries. However, before exoskeletons can become commonplace in industry as personal protective devices much like safety glasses or hardhats, researchers agree (Baldassarre et al., 2022; Elprama et al., 2020, 2022; Ferraro et al., 2020; Kermavnar et al., 2021; Kim et al., 2019; Tarbit et al., 2022) that we must first fully understand, demonstrate, and document the benefits of using exoskeletons in industrial work environments, and overcome the challenges and constraints industries face in adopting and implementing exoskeletons (Kermavnar et al., 2021; Looze et al., 2015, 2021). In particular, because exoskeletons are wearable devices, their effectiveness in practice may be dictated not only by how well their functional features support task performance but also by how comfortable the exoskeleton is for the worker to wear for prolonged periods of time (Baldassarre et al., 2022; Kim et al., 2019; Tarbit et al., 2022). In turn, worker comfort can be influenced by aspects such as the degree of anthropometric fit (Looze et al., 2015; Pesenti et al., 2021).

To address the significant knowledge gaps that prevent successful full-scale industry implementation of exoskeletons, researchers have been conducting studies evaluating occupational exoskeletons in laboratory environments and, in a few cases, in field settings. These lab-based studies have focused on how well specific types of exoskeletons support occupational task performance through the evaluation of muscle activity and how usable and comfortable they are through user feedback about task performance. Evidence from studies assessing the degree of support these exoskeletons provide through the evaluation of muscle activity, and biomechanical force and stress modeling indicates that in activities such as lifting, walking and overhead work activities, both passive and active exoskeletons reduce muscle activity by 20% to 80% (Baltrusch et al., 2018, 2019; Bock et al., 2022, 2023; Cha et al., 2019; Gillette & Stephenson, 2019; Harant et al., 2023; Hwang et al., 2021; Jackson & Collins, 2015; Kermavnar et al., 2021; Latella et al., 2022; Looze et al., 2015; Luger et al., 2021; Ogunseiju et al., n.d.; Pinho & Forner-Cordero, 2022; Qu et al., 2021; Schmalz et al., 2019, 2022; Steele et al., 2017; Walter et al., 2023). Reduced muscle activity has been observed in the lower and upper back, erector spinae, and shoulder and knee muscles. Some studies have shown that exoskeletons may decrease the metabolic costs of work (Baltrusch et al., 2019; Bock et al., 2022; Hwang et al., 2021; Walter et al., 2023). A few studies have reported less favorable muscle mechanics and even increases in muscle activation (Farris et al., 2014).

Many studies that quantify joint kinematics and muscle activity have also reported on participants' comfort, fit, and movement restrictions, all of which model the wearability of the exoskeleton and the overall helpfulness of the exoskeleton in assisting with the task (Chae et al., 2021; Dijsseldonk et al., 2020; Kim et al., 2018, 2019; Luger et al., 2021; Meyer et al., 2019). Participants reported increased discomfort when wearing the exoskeletons for tasks that required extreme postures, with individual fit and preferences influencing discomfort ratings. Furthermore, sex-related differences in how exoskeletons fit specific anthropometric profiles seem to play a role in discomfort ratings (Kim et al., 2020). Other similar studies have reported that participants experience discomfort around the knees and suggest thicker knee pads when using a lower-extremity exoskeleton (Abdoli-E et al., 2006). Additionally, studies have reported mixed discomfort scores for back support and trunk exoskeletons (typically compared with and without an exoskeleton) of the lower back and trunk (Antwi-Afari et al., 2021; Cardoso et al., 2020; Goršič et al., 2021; Kozinc et al., 2021; Madinei et al., 2020) and the forearms, upper arms, and shoulders (Daratany & Taveira, 2020; De Bock et al., 2020; Ferreira et al., 2020; Kim et al., 2018; Moyon et al., 2018; Van Engelhoven et al., 2018). Only recently have studies begun to focus exclusively on evaluating usability, rather than combining usability or discomfort evaluations as part of evaluating the functionality of exoskeletons. For example, in a recent study evaluating the usability of harnesses supporting a sit-stand exoskeleton, researchers assessed the usability in terms of wearability, stability, convenience, and overall wearing satisfaction (Chae et al., 2021). Wearability was assessed based on how easy it was to remember the steps to use, fasten, and adjust the harnesses. They assessed stability using wearing pressure, strength, comfort, safety, and overall stability with Likert-type rating scales. The convenience of the harnesses was assessed using thermal sensations, wetness sensations, and cushioning, as well as what participants felt about overall convenience. The authors concluded that stability and wearability significantly affected overall wearing satisfaction. Assessing usability with scales such as the System Usability Scale (SUS), with higher scores on the SUS indicating higher usability (Orekhov et al., 2021), also indicates a 45% increase in SUS scores when exoskeletons have been designed using user-centered design approaches (Meyer et al., 2019). While many of these studies have provided critical insights into the effectiveness of exoskeletons and on their usability and comfort during use in simulated or actual occupational tasks, we still do not know whether these exoskeletons are designed for universal use to accommodate a wide range of worker characteristics.

There is a need for exoskeletons, wearable devices, and PPE to be universally available and usable, based on the most recent labor force participation data (March 2024) published by the Office of Disability Employment Policy of the BLS. These data indicate that the labor force participation rate for persons with a disability aged 16–64 years is 40.3%. It is also well known that the US workforce has been aging and has grown by 117% in the last 20 years. More recent projections show that adults aged 65 and older will account for 9% of the labor force in 2032 and are expected to account for 60% of labor force growth from 2022 to 2032. Furthermore, many older workers are employed full-time rather than part-time. Recent data from the Current Population Survey and the BLS database indicate that nearly 45% of the labor force in 2032 is expected to be aged 45 and above, with only 44% between 25 and 44 years of age (BLS, 2023). Women aged 25–54 years continue to constitute a large percentage of the labor force, with 76.4% of women participating in the labor force in 2022. Women aged 55 years and older constitute approximately 33.6% of the labor force in 2022, and this number is expected to remain at 33% in the next 10 years (BLS, 2023).

The unique characteristics of the labor market and associated worker characteristics, and the potential need for these workers to use exoskeletons effectively as personal protective equipment in the future, necessitate an evaluation of whether and to what extent occupational exoskeletons are designed for *all*. When exoskeletons are designed for inclusivity and accessibility, all workers, including those with disabilities, can benefit from the potential for injury reduction that exoskeletons can offer by removing barriers to full participation by disabled workers without the need to significantly adapt these wearable devices to these workers. Furthermore, older workers and women in the workforce have different capabilities and limitations in anthropometry and fit, strength, and dexterity, which must be considered when designing exoskeletons. To date, no studies have evaluated whether these exoskeletons are designed for use by all.

Furthermore, most studies on usability and comfort have exclusively focused on evaluating exoskeletons when performing simulated occupational tasks. This is undoubtedly useful knowledge to acquire, but an exoskeleton as a wearable device, which is also modular in product design architecture and consists of loose parts, needs to be first assembled and then donned by the worker before the worker can use it for performing a task; after task performance is completed, the worker must then doff, disassemble and sanitize the exoskeleton before next use. An important rationale for considering these four tasks is how these basic tasks might represent what would typically occur in practice if exoskeletons were implemented in industry. This rationale dictates that exoskeletons be universally usable not only during task use, but also during assembly, donning, doffing, and disassembly. Currently, there is no knowledge on how the design of exoskeletons supports universal use during the assembly, donning, doffing, and disassembly stages; without assembly or donning, an exoskeleton cannot be used and evaluating exoskeletons during assembly, donning, doffing, and disassembly may reveal design problems different from, and not indicated by, evaluations conducted during task and use conditions. If these critical tasks are difficult to perform, industry adoption of exoskeletons could become limited, given the productivity and cost concerns industry will have in quickly being able to put these devices together, wear them, and remove them when needed.

To address these gaps, our aim is to understand the extent to which exoskeletons are designed for universal use *by all* workers. To achieve this aim, as one of the critical first steps, we conducted a heuristic evaluation of four exoskeletons using universal design principles as heuristic guidelines. The four different exoskeletons used in the study included: (1) a sit-stand exoskeleton used for alternating between sitting, standing and walking postures; (2) a hand grip strength exoskeleton; (3) a back support exoskeleton used to support lumbar flexion; and (4) a shoulder support exoskeleton used to support above shoulder work. Our evaluation also focused on four tasks that one would expect all exoskeleton users to perform to some degree when using the devices: assembling the exoskeleton, donning it, doffing it, and disassembling and storing it. The main study questions were as follows.

1. How well are exoskeletons designed for universal use? To answer this question, we evaluated four exoskeletons using universal design principles as heuristics: seven evaluators rated each exoskeleton to determine which principles the exoskeletons incorporated well into their designs, which universal design principles were violated, to what extent, and whether there were commonalities in violations of the design principles and design problems across exoskeletons.

2. Which tasks among assembly, donning, doffing, and disassembly of exoskeletons are easy to perform and which tasks are difficult to perform when evaluated against the principles of universal design? To answer this question, each evaluator tracked the violations of the universal design principles when performing these tasks with each exoskeleton. Evaluator ratings were then used to compare these task phases to determine specific design problems and heuristic violations in the assembly, donning, doffing, and disassembly tasks.

## 2. Methods

### 2.1. Study Approach

Our overall study approach consisted of three broad phases: an initial study planning and evaluator orientation phase, an evaluation and rating phase, and a rating discussion and reconciliation phase. The process steps are illustrated in Figure 1. These three project phases took approximately 4 months.

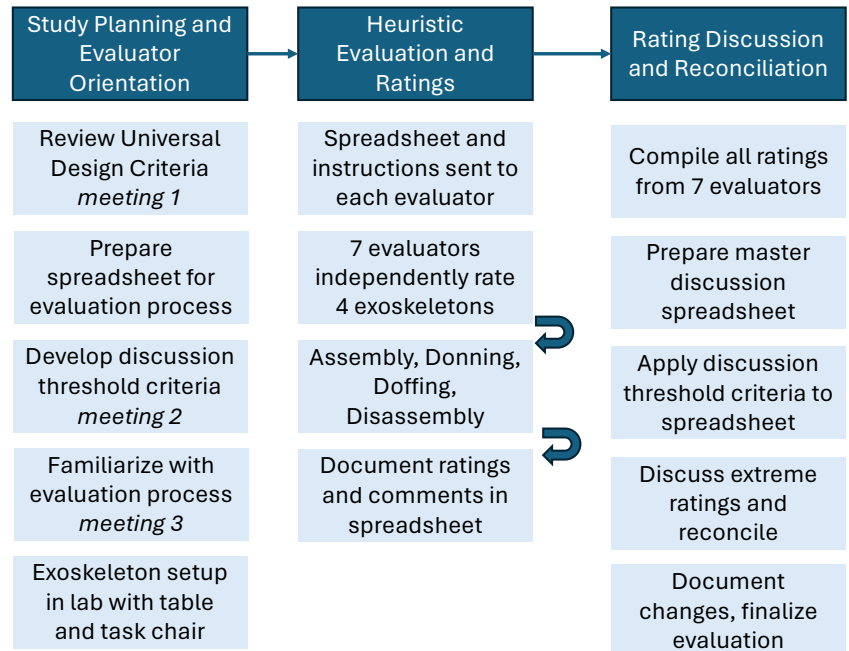

**Figure 1.** Three stages of heuristic evaluations with the major steps to proceed from planning to final evaluation documentation.

## 2.2. Exoskeletons Evaluated in the Study

In this study, we evaluated four commercially available exoskeletons. Each of these exoskeletons targets a specific movement and set of muscles that cause injury in the workplace. The exoskeletons evaluated include a lumbar flexion support device; a shoulder support device that is intended to reduce fatigue when workers perform activities above shoulder height; an exoskeleton for tasks that may alternate between sitting, standing, and walking postures, all three of which are passive exoskeletons; and a hand grip strength exoskeleton, which is an active exoskeleton powered by a battery pack.

## 2.3. Universal Design Principles and Evaluation Criteria Used

In our study, we used seven commonly known principles of universal design (Center for Universal Design, 1997), and the guidelines provided under each principle. We converted each guideline under each principle into a question format instead of leaving it in its prescriptive format so that evaluators could find it easier to anchor their evaluations. Table 1 lists the universal design principles and the evaluation criteria used in this study. Our rating scale was a Likert-type scale from 1 to 5, where a rating of 1 represented not at all, a rating of 3 represented moderate, and a rating of 5 represented an excellent design with respect to the universal design criteria for the task.

**Table 1.** Universal design principles and evaluation criteria used in the study. The numbers for the evaluation criteria correspond to the numbers in Figure 4 for the evaluation criteria on the x-axis.

| Design principle | Evaluation criteria (To what extent) |
|---|---|
| Equitable use | 1. Does the design provide the same means of use for everyone?<br>2. Does the design ensure that it is equally safe, private and secure for all?<br>3. Is the design aesthetic and unobtrusive?<br>4. Does the design avoid stigmatizing users<br>5. Does the design minimize memory load? |
| Flexibility in use | 1. Does the design provide adjustable features for diverse users? Consider male, female, old, young, different types of workers, demographics, etc.<br>2. Does the design support efficiency by providing flexibility during use?<br>3. Does the design consider portability of the device?<br>4. Does the design ensure adaptability of the product to the user's pace?<br>5. Does the design facilitate users' accuracy and precision? |
| Simple and intuitive to use | 1. Does the design support learnability and intuitiveness?<br>2. Does the design eliminate unnecessary complexity?<br>3. Does the design consider a wide range of language and literacy skills?<br>4. Does the design ensure consistency of functional features?<br>5. Does the design assure match between system and world?<br>6. Does the design ensure that users are in control?<br>7. Does the design make system state visible?<br>8. Does the design provide clear closure in task? |
| Perceptible information | 1. Does the design consider different modes of information? |

| Design principle | Evaluation criteria (To what extent) |
|---|---|
| | 2. Does the design users' sensory limitations? |
| | 3. Is the design minimalist? |
| | 4. Does the design use users' language? |
| | 5. Does the design provide informative feedback? |
| | 6. Does the design provide help when needed? |
| | 7. Does the design provide good error messages? |
| | 8. Does the design maximize legibility of essential information? |
| Tolerance for error | 1. Does the design prevent errors before it occurs? |
| | 2. Does the design protect users from unintended misuses? |
| | 3. Does the design support conscious actions that require vigilance? |
| | 4. Does the design provide necessary warnings? |
| | 5. Does the design arrange elements in a way that minimizes hazards or errors? |
| | 6. Does the design ensure durability? |
| | 7. Does the design provide help and support for recovering from problems? |
| | 8. Does the design ensure that actions are reversible? |
| Low physical effort | 1. Does the design maintain a neutral body posture? |
| | 2. Does the design ensure that the operating forces are reasonable? |
| | 3. Does the design minimize repetitive actions? |
| | 4. Does the design minimize sustained physical effort? |
| Size and space for approach and use | 1. Does the design provide a clear line of sight to important elements for any seated or standing user? |
| | 2. Does the design make sure that all components are comfortable to reach for any seated or standing user? |
| | 3. Does the design provide adequate space for use of assistive devices or personal assistance? |
| | 4. Does the design accommodate variations in hand and grip sizes? |

*2.4. Tasks Evaluated in the Study*

We evaluated four tasks: assembly of the exoskeleton, its donning, doffing, and its disassembly. Each exoskeleton came from the manufacturer disassembled into its component parts either in a manufacturer-supplied bag, a case, or a box.

For the purpose of the evaluation, we defined an assembly task as being able to assemble all the parts from the exoskeleton storage bag/box into a complete exoskeleton that is ready to be put on. The donning task consisted of successfully putting on the exoskeleton to the point at which the exoskeleton was ready for use. Doffing was defined as the successful removal of the exoskeleton from the body and its placement on a desk or work surface. The disassembly task consisted of removing and separating all the parts of the exoskeleton so that it would be ready for assembly the next time. For all four tasks, the evaluators could use the instruction manuals that either came in the box or were available online and hand tools such as the hex keys that came with the exoskeletons.

*2.5. Evaluation process and procedure*

The evaluation process is illustrated in Figure 1 and the universal design guidelines are listed in Table 1. After planning and orientation, the evaluators agreed to independently evaluate the four exoskeletons and complete a spreadsheet with detailed comments on the problems encountered when assembling, donning, doffing, and disassembling each exoskeleton with respect to each evaluation criterion (see Figure 2).

**Figure 2.** Spreadsheet template used for completing the evaluations.

The evaluators were instructed to use a worktable (available in the laboratory) as the main surface to place the exoskeletons for any assembly and disassembly so that we could maintain uniformity with respect to the postures, heights, distances, and reaches encountered during the evaluations. The evaluators were asked to perform any required sitting posture during the evaluation using a standard task chair available in the laboratory.

The consolidated master spreadsheet template for combining all individual files from all evaluators is shown in Figure 3. The master sheet contained columns for the task (assembly, donning, doffing, and disassembly), design principles, evaluation questions, and rating columns for each of the seven evaluators with their completed ratings and their comments justifying their ratings.

[Figure 3 spreadsheet screenshot showing columns: Task, Design Principle, Evaluation Criteria, Rating, Comment (repeated for multiple evaluators). Rows under Task "Assembly" and Design Principle "Equitable use" include criteria: "Does the design provide the same means of use for everyone?", "Does the design ensure that it is equally safe, private and secure for all?", "Is the design aesthetic and unobtrusive?", "Does the design avoid stigmatizing users?", with ratings and comments from multiple evaluators.]

**Figure 3.** Screenshot of compiled master spreadsheet for one exoskeleton. Each exoskeleton had its own worksheet in the workbook, which contained ratings and comments for all four exoskeletons.

## *2.6. Rating Discussion and Reconciliation Process*

The final phase of the evaluation process involved a discussion of the ratings assigned by the evaluators and a reconciliation and adjustment of the ratings if the ratings diverged extensively among the evaluators. To prepare for discussions to reconcile the ratings, the team prepared a list of rules to decide whether a certain set of ratings for a criterion needed further discussion and reconciliation. The most important rule was to examine the extreme ratings from the evaluation and determine which criteria needed further discussion. A score of 1 on the rating scale indicated a poor design and a score of 5 on the rating scale indicated an excellent design with respect to the universal design criteria. The process for determining the need for discussion is illustrated in Figure 4.

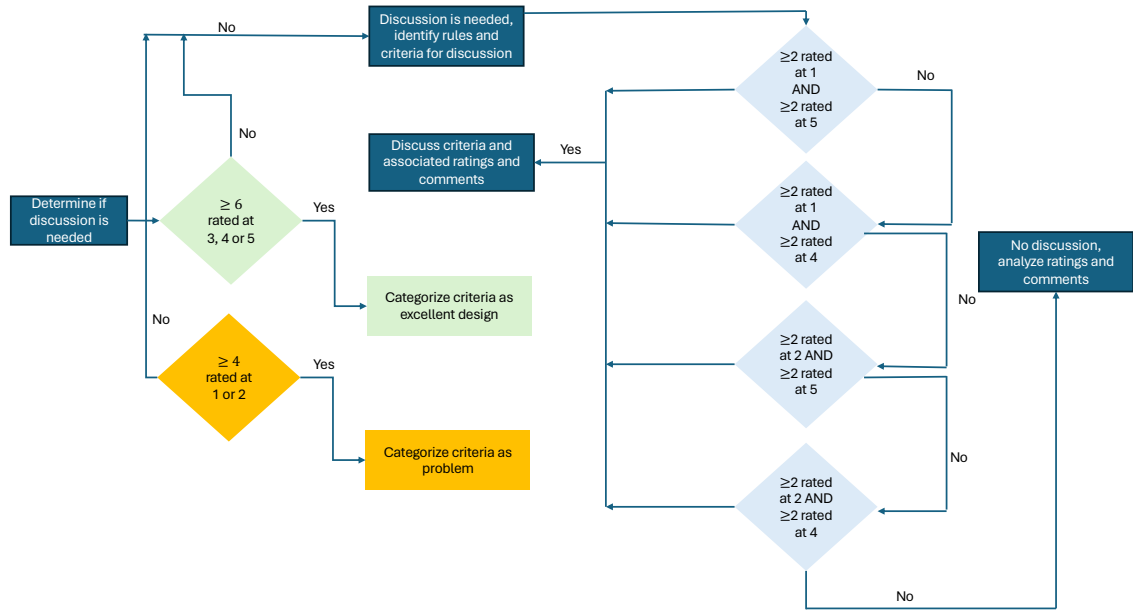

**Figure 4.** Decision process for determining the need for discussion of ratings using threshold criteria and rules.

Our logic for deciding the threshold rating levels for what needed discussion and what did not was that even a little evidence may be enough to warrant discussion for a design problem; hence, a threshold of two raters for discussing ratings that exhibited spread and that we would demand more evidence (a greater number of raters) to agree that the design was good to excellent with respect to universal design criteria; hence, the threshold of six raters or all seven raters for ratings of 3, 4, or 5. During the discussion sessions, the team first heard from the extreme raters regarding the rationale for their numerical rating for each task and for each exoskeleton. Other team members then joined in the discussion. Based on this discussion, team members either elected to stay with their numerical rating for a criterion or decided to change their rating. In general, three reasons emerged for the disagreements in the ratings: (1) some evaluators felt that the assembly and the donning tasks could not be clearly separated during the evaluation because some assembly occurred after the evaluator had donned the exoskeleton; (2) some evaluators felt that the device itself had to provide information such as feedback to the user and not have the user refer to the manual (which in most cases had the information in printed form); and (3) some evaluators felt that if the user was able to complete a task, such as assembly or donning, that they would also be able to complete the complimentary task such as disassembly and doffing, and therefore provided a similar rating. Table 2 shows the number of criteria that warranted discussion and were resolved after discussion.

**Table 2.** The number of criteria that were resolved after the discussion. We applied the same threshold rater-rating rules used prior to the discussion to decide whether the discussions resolved the disagreements in the ratings.

| Type of Exoskeleton | Number of criteria discussed | Number of unresolved and unreconciled criteria after discussions |
| --- | --- | --- |
| Back support exoskeleton | 20 | 2 |
| Shoulder support exoskeleton | 31 | 3 |
| Grip strength support exoskeleton | 6 | 0 |
| Sit-stand support exoskeleton | 26 | 4 |

### *2.7. Statistical analyses*

We generated descriptive statistics, including means, frequencies, and counts, as follows.

a. The mean ratings for universal design principles that received a rating of 3 or below were averaged over all the criteria for a principle and averaged over all seven raters and all four tasks (assembly, donning, doffing, and disassembly).

b. The number of times a universal design criterion received an average rating of three or below when averaged over the four tasks and four exoskeletons.

c. a ranking of the top three criteria that evaluators rated at 1 and 2 sorted by the number of raters assigning the ratings and categorized by the task and design principle for each of the four exoskeletons.

d. categorization of the number of violations of universal design criteria across all four exoskeletons, organized by task stage (assembly, donning, doffing, and disassembly).

All analyses were tabulated and/or appropriate visualizations were generated.

**3. Results**

Our research questions were twofold: How well do exoskeleton designs, in general, adhere to the principles of universal design, and which tasks among assembly, donning, doffing, and disassembly of exoskeletons are easy to perform, and which tasks are difficult with respect to universal design principles?

The scores in Table 3 indicate that the universal design principles of providing perceptible information to the user, making the design equitable to use, and making the design simple and intuitive to use were design principles on which the evaluators rated the exoskeletons poorly. Furthermore, these design principles received average ratings of less than three for assembly and donning tasks; disassembly tasks were rated at an average of three or above, indicating that disassembly did not significantly violate universal design principles. It is also noteworthy that the hand grip strength exoskeleton did not receive any average ratings at or below three for any of the design principles.

**Table 3.** Design principles that received an average rating of less than 3 when averaged over all criteria for that principle over all raters for different tasks.

| Exoskeleton | Task | Design Principle | Average Score |
|---|---|---|---|
| Sit-stand | Assembly | Perceptible information | 2.77 |
| | Donning | Equitable use | 2.91 |
| Back support | Assembly | Equitable use | 2.71 |
| | Assembly | Simple and intuitive | 2.84 |
| | Assembly | Perceptible information | 2.54 |
| Shoulder support | Donning | Size and space | 2.95 |
| | Assembly | Equitable use | 2.94 |
| | Assembly | Simple and intuitive | 2.85 |
| | Assembly | Perceptible information | 2.96 |
| | Donning | Equitable use | 2.88 |

The results indicate that the principles of providing perceptible information to the user, making the design equitable for use by all, and designs incorporating tolerance for user errors were the top three principles that were all poorly rated at 1 and 2 by a majority of raters (Table 4). The specific criteria varied from a lack of safety features in some exoskeletons, such as the sit-stand and shoulder support exoskeletons, to a lack of informative feedback for the user from some exoskeletons, such as handgrip strength and sit-stand exoskeletons (Table 4 and Figure 5). Assembly and donning tasks represented the majority of the violations of the criteria.

**Table 4.** The top 3 universal design evaluation criteria that obtained the most scores of 1 and 2 from the evaluators. The criteria are tagged by the task phase and the design principle violated for each exoskeleton.

| Exoskeleton | Task | Design Principle | Universal Design Criteria | Number of evaluators rating at 1 and 2 |
|---|---|---|---|---|
| Sit-stand | Donning | Equitable use | Equally safe, private and secure for all | 7 |
| | Donning | Tolerance for error | Supports conscious actions that require vigilance | 7 |
| | Assembly | Perceptible information | Provide informative feedback | 6 |
| Handgrip strength | Doffing | Perceptible information | Provide informative feedback | 6 |
| | Donning | Perceptible information | Provide informative feedback | 6 |
| | Assembly | Tolerance for error | Provide necessary warnings | 4 |
| Back support | Assembly | Perceptible information | Maximize legibility of essential information | 6 |
| | | | Minimalist design | 6 |
| | | | Consider users' sensory limitations | 6 |

| Shoulder support | Assembly | Simple and intuitive | Consider a wide range of language and literacy skills | 5 |
| --- | --- | --- | --- | --- |
| | | | Ensure consistency of functional features | 5 |
| | Donning | Equitable use | Equally safe, private and secure for all | 5 |

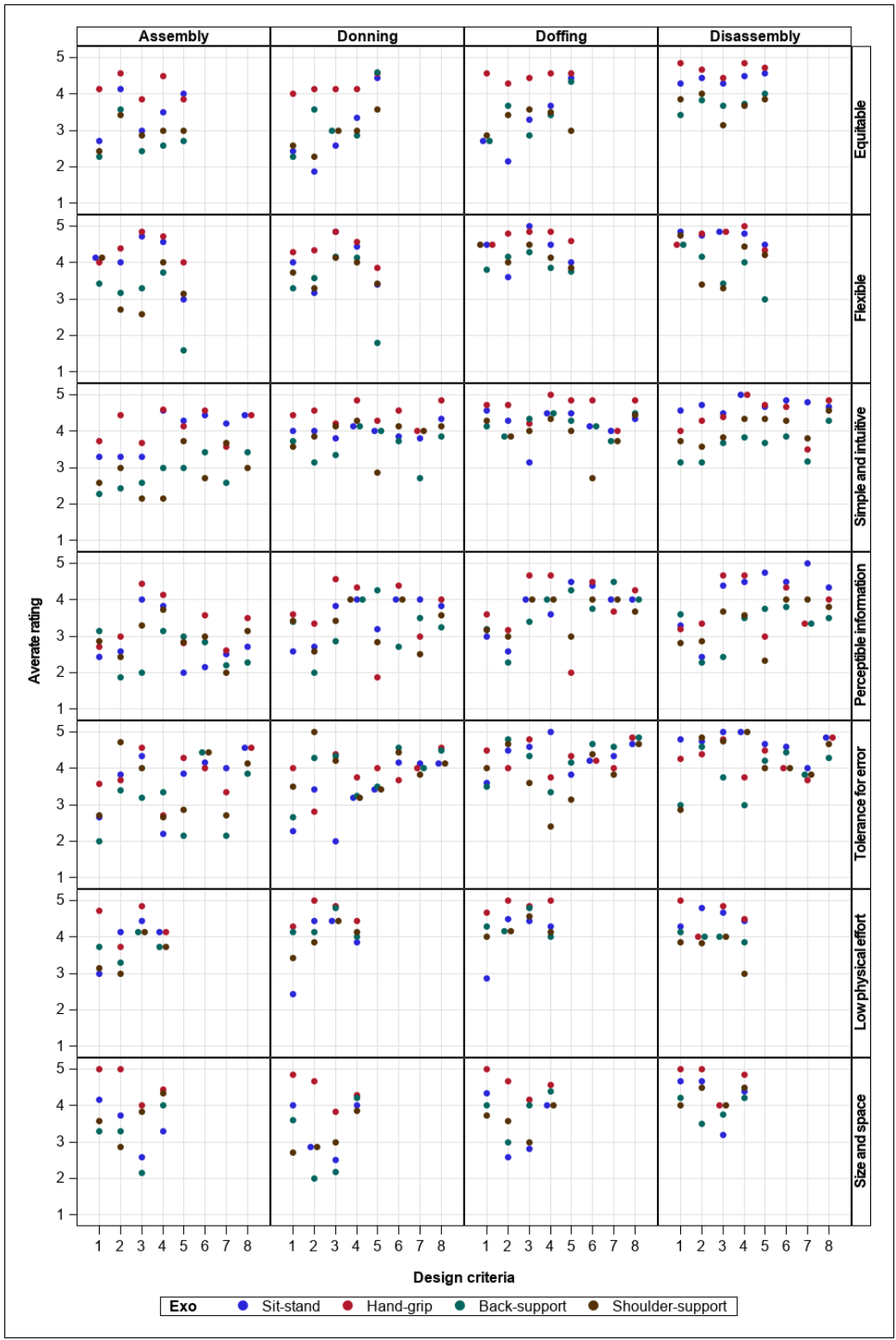

**Figure 5.** Average rating for all 7 universal design principles for each evaluation criterion (numbered in Table 1) for each exoskeleton included in the study by each task phase including assembly donning, doffing, and disassembly and storage.

Table 5 indicates that the design criterion that had the most violations with a rating of 3 or below was failing to consider user's sensory limitations.

**Table 5.** Frequency of violations of universal design criteria that received an average rating of 3 or below for all tasks and exoskeletons.

| Universal design criteria | Frequency of violations |
|---|---|
| Consider users' sensory limitations | 13 |
| Provide informative feedback | 10 |
| Provide same means of use for everyone | 9 |
| Provide adequate space for use of assistive devices | 7 |
| Be aesthetic and unobtrusive | 7 |
| Prevent errors | 7 |
| Provide good error messages | 6 |
| Consider different modes of information | 6 |
| Comfortable to reach in seated or standing postures | 6 |
| Provide necessary warnings | 5 |

Not all universal design principles and criteria fared poorly. All seven evaluators rated aspects such as not having to remember what actions to take; the design requiring only minimal repetitive actions; users having the possibility of reversing their actions quickly; the designs of the exoskeletons affording assembly, donning, doffing, and disassembly tasks at the user's pace of work; and reasonable force and strength demands, among others, as positive features (Table 6).

**Table 6.** Universal design criteria, corresponding principles and the tasks that were rated between 3 and 5 by all 7 evaluators for each exoskeleton.

| Exoskeleton | Task | Design Principle | Universal design criteria |
|---|---|---|---|
| Sit-stand | Assembly | Equitable use | Minimize memory load |
| | Donning | Low physical effort | Minimize repetitive actions |
| | Doffing | Simple and intuitive | Support learnability and intuitiveness |
| Hand grip | Assembly | Flexibility in use | Ensure adaptability of product to user's pace |
| | Donning | Tolerance for error | Ensure actions are reversible |

|   |   | Equitable use | Be aesthetic and unobtrusive |
|---|---|---|---|
| Back support | Donning | Low physical effort | Ensure operating forces are reasonable |
|   |   |   | Minimize sustained physical effort |
|   |   | Simple and intuitive | Ensure users are in control |
| Shoulder support | Donning | Flexibility in use | Provide adjustable features for diverse users |
|   | Donning | Simple and intuitive | Ensure that users are in control |
|   | Disassembly | Simple and intuitive | Provide clear closure in task |

Figure 6 illustrates the number of violations across all exoskeletons by task phase, indicating that the assembly and donning of the exoskeletons violated more universal design criteria than the doffing and disassembly tasks. This trend is also reflected in Tables 3 and 4, where most of the violations of the design principles and criteria occur during the assembly and donning phases.

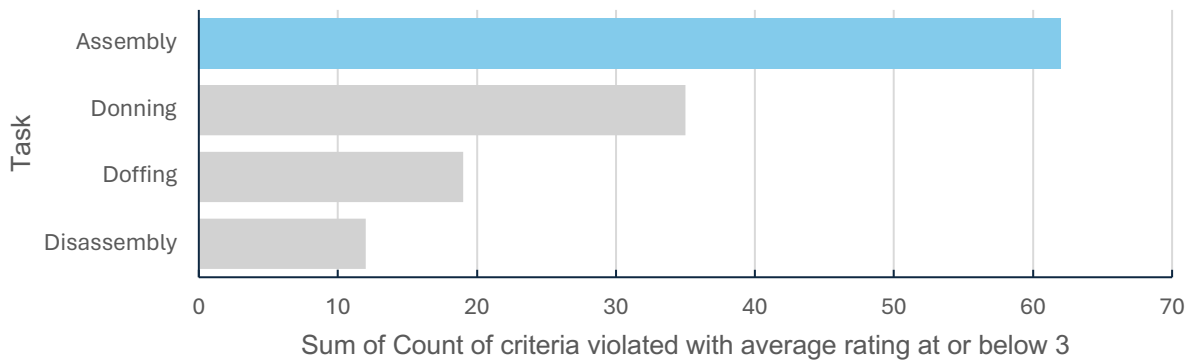

**Figure 6.** The number of criteria violations of the universal design heuristics for assembly, donning, doffing and disassembly tasks across all exoskeletons. Assembly and donning tasks accounted for 76% of the 128 violations.

## 4. Discussion

Newer occupational exoskeletons are continually being evaluated by researchers for their efficacy in directly supporting worker performance in tasks that may cause musculoskeletal injuries. However, from a design standpoint, we do not know how well exoskeletons, in general, are designed for universal use by *all* workers, which is particularly important given the diversity in the industrial workforce, including workers with disabilities, workers with work-related injuries or mobility challenges, older workers, and female workers. Furthermore, it is unclear how well exoskeleton designs afford the tasks of assembly, donning, doffing, and disassembly; if the exoskeletons cannot be quickly assembled, donned, and doffed when needed, industry is unlikely to embrace and adopt them given the productivity concerns industry is likely to raise. To answer these two research questions, in our study, we heuristically evaluated four different occupational exoskeletons using widely accepted universal design principles and criteria when assembling, donning, doffing and disassembling the devices. We further discuss our findings on how well the four exoskeletons fared when evaluated against the universal design principles.

### *4.1. Universal design principles that were rated poor in the exoskeletons*
### *4.1.1. Design for equitable use*

The design of the exoskeletons for equitable use, as a universal design principle, received an average rating of less than 3 across three exoskeletons: the sit-stand, back support, and shoulder support exoskeletons (Table 3). In particular, in the sit-stand and shoulder support exoskeletons (Figure 5), all seven evaluators perceived that the exoskeletons were not equally safe for all users during donning and rated them only in the range of 1 to 2 (Table 4). We believe this is the case because the sit-stand exoskeleton requires the wearer to balance or to have another person help them while donning the exoskeleton and when attempting to sit on the seat pads to avoid the risk of falling. Bending actions are required to secure the exoskeleton to the feet, introducing an additional risk of falling owing to the weight of the exoskeleton on the body during bending. In addition, if users assemble the exoskeleton while donning it, there is an even greater risk of falling. When donning the sit-stand exoskeleton, the user must hold the device while securing it to the body because the device does not stand upright without support. Similarly, with the shoulder support exoskeleton, multiple evaluators commented on the safety risks of accidentally hitting themselves with the unanticipated bouncing back and ricochet of the arm-cup assembly when donning the exoskeleton. These are significant safety concerns that warrant these severity ratings.

Our findings indicate that potential safety concerns when donning exoskeletons have not received sufficient attention in the design of these devices. We believe that one reason could be the importance provided during design to use these exoskeletons for task performance, compared to assembly or donning activities. Additionally, current evaluations of usability have focused primarily on use conditions and task simulations. Our study identifies a significant knowledge gap in the recognition of these safety concerns. Furthermore, these concerns were identified only because we evaluated the assembly, donning, doffing, and disassembly phases, pointing to the need for a comprehensive evaluation of exoskeletons during the pre- and post-use phases, and not only during use for task performance.

Together, the violations in the equitable design criteria suggest that the evaluated exoskeletons do not adequately consider all the user populations and characteristics. Both safety concerns can be exacerbated in people with balance problems or those with limited mobility. These findings indicate that exoskeletons are currently not usable by workers with diverse abilities, or users with canes, or other assistive devices. It is not clear what user population was used as a baseline for designing exoskeletons. Recent data show that the labor force participation of people with disabilities in the working age range of 16–64 years is nearly 40%. One could argue that these workers would benefit more from these exoskeletons. Therefore, these worker characteristics must be considered when designing an exoskeleton.

### 4.1.2. Perceptible information from the design

Perceptible information from the design for the user, as a principle, considering all its criteria, received an average rating of less than 3 across three exoskeletons: the sit-stand, the back support, and the shoulder support devices (Table 3). Two exoskeletons, the sit-stand device and the handgrip strength device (Figure 5), received ratings of 1 to 2 from six evaluators for not providing informative feedback across three different tasks: assembly, donning, and doffing (Table 4). Additionally, informative feedback criteria were the second most frequent, with average ratings of 3 or below across all exoskeletons and tasks (Table 3). There was considerable discussion among raters regarding whether and how exoskeletons, as passive devices, should provide feedback during assembly, donning, doffing, and disassembly. The evaluators agreed that the design itself should provide feedback when attaching and detaching parts and that the device design did not provide sufficient feedback for users to move through a sequence of steps to facilitate assembly. In the case of the active hand grip exoskeleton, the evaluators felt that it was a missed opportunity to not provide feedback on the device itself during donning and doffing, particularly when the user actions were incorrect.

Six evaluators provided scores in the range of 1 to 2 for three criteria related to perceptible information (Table 4) for the back-support exoskeleton: (1) legibility of information, (2) minimalist design, and (3) sensory limitations. First, the evaluators believed that the sizing information was not legible on the exoskeleton. The sizing was color-coded, but it was a thin part of the exoskeleton and not very apparent to the user. In addition, users with visual impairments may have difficulty differentiating between these colors. Second, no information was available on the sequence of assembly on the device itself. For example, there is no information on how the vest holding the exoskeleton should be assembled together. Therefore, the manual was consulted to identify the correct sizing and assembly steps. However, consulting the manual did not help either - the information in the manual was too small and not legible.

The evaluators overwhelmingly felt that there were too many component parts in the exoskeleton, which made the assembly task exceedingly difficult and tedious. Furthermore, the assembly required special tools and a flat surface to hold multiple parts. Hence, evaluators felt that the design was not minimalist. We believe that one reason the design is not minimalist is the "toughness" in engineering the functions required of exoskeletons may have unfortunately also permeated to the looks, number of parts, and steps required to assemble the exoskeletons. This may make one feel that the assembly task is extremely complex. We believe that the burden of assembly and donning is an important determinant of industry adoption and worker compliance, emphasizing the need to evaluate these conditions during design and to make assembly and donning as seamless as possible.

The evaluators felt that the potential user must primarily rely on sight and auditory and tactile feedback from clicks to assemble the back-support exoskeleton. In particular, given the large number of parts required for assembly and the need to rely on a manual, the exoskeleton appears to rely on vision as the primary mode for the user to obtain information from the exoskeleton. For example, the exoskeleton contains key information such as sizing only in the visual mode using colors. Our analysis of the frequency of criteria violations also revealed that the sensory limitation criteria received average ratings at or below 3 more often than the other criteria across all exoskeletons and tasks (Table 5). Given these findings, those with sensory limitations will be unable to assemble and have a proper fit in these exoskeletons without assistance. Given that workers with disabilities have the potential to play an important role in industry, future exoskeleton designs need to consider universal design principles for users with diverse sensory abilities, especially if exoskeletons are to formally become personal protective equipment (PPE) in industry.

Criteria violations in the perceptible information principle suggest that designers place more emphasis on the functions that need to be supported by these exoskeletons compared to the form and user-exoskeleton interactions. Exoskeletons are meant to provide functional support to the relevant muscle groups that activate during work tasks (Baltrusch et al., 2018, 2019; Bock et al., 2022, 2023; Cha et al., 2019; Gillette & Stephenson, 2019; Harant et al., 2023; Hwang et al., 2021; Jackson & Collins, 2015; Kermavnar et al., 2021; Latella et al., 2022; Looze et al., 2015; Luger et al., 2021; Ogunseiju et al., n.d.; Pinho & Forner-Cordero, 2022). While providing superior functional support is an important consideration and the primary reason for wearing an exoskeleton, we believe that user-exoskeleton interaction elements need more attention, especially during the assembly, donning, doffing, and disassembly of exoskeletons. If ease of assembly and donning is a trade-off for superior functions, long-term user compliance may suffer.

We also believe that many criteria under the perceptible design principle were violated, perhaps because of the relationships between the size, shape, and surface area of the exoskeletons and how best information and instructions can be presented on the device itself for assembly and donning.  In general, exoskeletons do not have a large surface area for displaying information crucial for assembly, donning, doffing, and disassembly. Many parts of an exoskeleton are thin, tubular, curved, with little surface area, and have hidden slots to fit the curvature of the body, thereby limiting opportunities to display external help or instructions on the device itself.

Three of the four exoskeletons that were evaluated must be worn on a person's back. This presents additional design challenges for making the information legible and perceptible, particularly when donning the device. In some exoskeletons, we found that assembly and donning overlap, making visual feedback difficult for the user. Additionally, the fit is determined by a visual check once the exoskeleton is donned and then by making finer adjustments. The user does not have any direct visual feedback when donning the exoskeleton; therefore, adjustments are made through trial and error. Given the lack of visual access and feedback, a second person may be required to assist with the donning of exoskeletons for a proper fit. Future designs should consider how best to present information about assembly, donning, doffing, and disassembly to users to reduce user burden and improve the usability and accessibility of user-exoskeleton interaction points.

### 4.1.3. Tolerance for error

Preventing errors received an average rating of 3 or fewer, seven times across all exoskeletons and tasks (Table 5). The active hand grip exoskeleton received ratings of 1-2 from four evaluators (Table 4) owing to the lack of visual, auditory, or other modes of warning on the device itself. The hand-grip exoskeleton has sensors that are activated when connected to a battery. The sensors can be damaged if the hand-grip glove is not assembled and connected properly to the battery. The manual warns about these aspects, but the user must consult it to learn them. Given that this exoskeleton was active, the evaluators felt that the device itself could incorporate and provide visual or auditory warnings to the user.

All evaluators provided ratings of 1–2 for the vigilance criteria for the sit-stand exoskeleton (Table 4). The evaluators had considerable discussion on this criterion; they felt that the sit-stand exoskeleton required too much vigilance to don carefully without losing balance or falling. Although the criteria evaluated conscious actions that require vigilance as a positive characteristic, the evaluators felt that because of safety concerns, requiring constant vigilance from the user was a negative design characteristic, forcing us to rate the criteria lower to emphasize fall risk as an important safety concern.

The violation of these criteria under the tolerance-for-error design principle emphasizes the safety hazards from the sit-stand exoskeleton and concerns related to the maintenance and durability of the exoskeleton. The tolerance for error principle ensures that hazards are minimized for all people. For example, if a user with visual impairments or a user with mobility and balance issues were to use the sit-stand exoskeleton, it is not clear how they would perform the conscious and mostly visual actions with vigilance, without the support of another person, to ensure they do not fall and that they wear the exoskeleton correctly. We believe that different use cases and user populations need to be adequately considered during design so that tolerance for errors can be built into the design.

**4.2. Universal design principles and criteria that were rated high in exoskeletons**

The evaluators also provided scores in the range of 4 to 5 for some criteria and tasks across the different exoskeletons. Across the three different criteria under the low physical effort design principle and across the two exoskeletons (Figure 5), evaluators provided scores ranging from 4 to 5 (Table 6). It is worth noting that these 4 to 5 scores for low physical effort are during donning. The evaluators felt that, once assembled, some of these exoskeletons did not require significant physical effort to wear it. These findings indicate that across exoskeletons, the physical effort required when donning the exoskeleton is minimal. The evaluators also felt that the design of the sit-stand exoskeleton minimized repetitive actions while donning the exoskeleton. In addition, the operating forces required to don the back-support exoskeleton were considered reasonable. This is beneficial for users with limited mobility or diverse abilities to don the exoskeleton without significant challenges.

The evaluators rated two criteria under the equitable design principle with an average rating of 4 to 5 across two different exoskeletons and tasks (Figure 5; Table 6). They felt that the sit-stand exoskeleton did not require memorization of the steps during assembly. Given that the evaluators followed the assembly steps from the manual and because the parts in the exoskeleton resembled the shape of the body parts, memorization was not required. Additionally, the hand-grip exoskeleton was rated as aesthetic and unobtrusive. We believe that this is because the hand-grip exoskeleton has the least number of parts and looks the most intuitive in form resembling a regular glove, making it more aesthetically pleasing and simpler to don. The evaluators felt that the exoskeleton would be enhanced if the glove could be worn cordlessly. The hand grip exoskeleton also received a score of 4 to 5 for the tolerance for error design principle (Table 6) because evaluators felt that the actions were easily reversible if the users made any errors and had to retrace their steps.

The evaluators felt that three different criteria under the simple and intuitive design principle across three different exoskeletons and tasks deserved scores of 4 to 5 (Table 6). The evaluators felt that doffing the sit-stand exoskeleton was simpler and intuitive once the user knew how to put it on. They also felt that the design of the exoskeletons allowed users to be in control. For the back-support exoskeleton, users have full control when donning the exoskeleton. For the shoulder support exoskeleton, although evaluators felt that they were in control when donning, the rebound of the arm posed control issues for the user. However, the evaluators thought that it was easy to disassemble the shoulder support exoskeleton, given that it could be taken apart in any sequence, providing clear closure during the task.

Two exoskeletons across two different tasks received scores in the range of 4 to 5 for the flexible design principle (Figure 5, Table 6) because of the adaptability and provision of adjustable features for diverse users. For example, for the hand-grip exoskeleton, evaluators thought that one could assemble the exoskeleton at their own pace, and there were no constraints dictating the user's pace. The shoulder support exoskeleton provides adjustable parts, making the entire design adjustable to different users. However, the evaluators recommended that the chest straps be redesigned with female users in mind.

### 4.3. Tasks that presented the most problems for evaluators

Our findings revealed that the assembly and donning stages presented the most problems for evaluators and consequently received the greatest number of average ratings at or below 3 (Figure 6). Some exoskeletons require time and effort for the user to develop a visual map of the many parts provided and to refer to the manual constantly to understand the sequence of steps required for assembly. Assembly also requires actions that demand force, strength, mobility, and balance to correctly assemble parts. Additionally, evaluators felt that some exoskeletons, such as the sit-stand device, contained significant overlaps between the assembly and donning tasks for the user to complete an effective sequence, but the instructions provided in the manual did not clarify this. These overlap points also posed safety risks; however, the design did not include any constraints to prevent the user from working around the instructions.

Our finding that approximately 75% of the violations were in the assembly or donning stages (Figure 6) is concerning given that one must progress through these stages to use the exoskeleton. This finding adds new knowledge about exoskeletons' use when assembling, donning, doffing, and disassembling them, and provides an impetus for designers and manufacturers to emphasize these stages in design so that users can successfully and easily complete these critical pre-use and post-use tasks. Future research should examine how factors related to industry adoption and worker compliance are affected by assembly, donning, doffing, and disassembly.

Although the doffing and disassembly stages did not pose as much concern as the assembly and donning stages, they posed some safety challenges. Similarly, some doffing actions require vigilance to prevent safety hazards. For example, removing the sit-stand exoskeleton requires bending to detach the feet that are connected to the entire exoskeleton still on the person. Similarly, detaching arm cuffs from the shoulder support exoskeleton sometimes results in the bounce-back and ricochet of a heavy metal arm, introducing a safety hazard. The disassembly task, like the assembly task, required strength to detach parts in some exoskeletons – sometimes they required using both hands to press down on levers to release the parts, not only requiring force, but also introducing pinch hazards. In most cases, doffing and disassembly required steps in the reverse order of assembly and donning, which explains the fewer violations of criteria in these two stages compared with assembly and donning.

Our overall findings across all exoskeletons suggest that current exoskeletons do not adequately consider all users and conditions in their design. Although some features and functions adhere to universal design principles, many actions required for assembly and donning do not account for all users. We recommend that future exoskeleton designs consider universal design principles such that all workers, regardless of their age, ability, functional status, and other characteristics, can access, use, and benefit from these exoskeletons. This is even more important if these exoskeletons are to be used as personal protective equipment in the future.

### 4.4. Design Implications

In general, our heuristic evaluation indicated six major implications for design: (1) evaluation of pre- and post-use tasks (assembly, don, doff, disassembly) during design; (2) consideration of user-exoskeleton interaction points; (3) consideration of a wider range of user population and characteristics when designing occupational exoskeletons; (4) address safety concerns; (5) highlight two-person operation or actions requiring assistance; and (6) consider work in situ factors.

First, exoskeleton designs should consider all tasks and actions necessary to use the exoskeleton, not just the use conditions or the functional performance support it provides to reduce injuries. We argue that it is important to consider the assembly, donning, doffing and disassembly of exoskeletons; without successfully completing these tasks, an exoskeleton cannot be used effectively. Exoskeletons are typically not a single device; they are composed of individual parts, which necessitate the user to either assemble them before use or have space and other resources to keep them assembled. In addition, if every worker does not receive an individual exoskeleton, the size must be adjusted, necessitating assembly. Although we believe that users will learn to assemble these devices over time, we suspect that the time and effort required to assemble them will not change significantly given the many parts these exoskeletons contain that must all be assembled. Most importantly, workers with diverse abilities may not be able to assemble or don these exoskeletons without additional support. The design and testing of exoskeletons need to evaluate how best to reduce the assembly burden of users, either by producing a partially assembled piece that is ready to wear or by changing the designs and constraints to make the assembly effortless and time-efficient for the user. Additionally, the conditions required for each of these tasks should be considered during the design. For example, if assembly is required every time, a flat surface may also be required. While the design itself should consider these pre- and post-use conditions and tasks, the assistance and documentation to facilitate these tasks should also be considered in conjunction with the exoskeleton design. We consider these as important determinants of industry adoption.

In particular, the design and testing phases should consider the user-exoskeleton interaction points during all the pre-use, use, and post-use conditions. These user-exoskeleton interaction points are key leveraging opportunities for making a user's experience with exoskeletons seamless and making the overall design more accessible and usable. For example, if the user needs to attach a back support frame to an articulating lever, the user–exoskeleton interaction point can use a locking mechanism that readily clicks into place, preventing the user from exerting a large force to attach the parts. When designing these user-exoskeleton interaction points, design constraints must also be considered so that the user does not have a chance to commit an error. Furthermore, during design, steps with potential overlaps between assembly and donning tasks that may impact the user-exoskeleton interaction points should be considered. It can be difficult for users to interact with an exoskeleton to attach, lock, or adjust parts given that many important exoskeletons are worn on the back or shoulders and do not provide sufficient visual or physical reach. These user-exoskeleton interaction points need to be evaluated not only during pre-use, use, and post-use conditions, but also with different user populations and characteristics in mind.

The design of exoskeletons should consider a wider range of user characteristics. Recent data show that workers with disabilities participate in large numbers in the workforce, so ensuring that they can also assemble, don, use, doff, and disassemble these exoskeletons is important. Furthermore, workers who may have already experienced injuries or who have medical conditions would benefit significantly from using exoskeletons for support during their work tasks. However, for this benefit to materialize, exoskeletons need to be designed to enable these workers to assemble, don, use, doff, and disassemble without exerting or injuring themselves further. Exoskeletons are extensively used in rehabilitation. Therefore, evidence from research in rehabilitation science may be useful for informing the design of occupational exoskeletons for all workers. Exoskeletons are also likely to be used by workers with diverse characteristics; therefore, design should consider users' literacy skills and educational backgrounds, particularly when designing help and documentation. To ensure that exoskeletons are designed for all, universal design principles can be used as guidelines for design and testing, in addition to usability evaluations and user testing to eliminate problems due to poor design.

      Potential safety concerns that arise during pre- and post-use conditions need to be addressed in the design. These safety concerns can pose a significant risk to any user, particularly those with mobility or balance problems. We believe that some of these safety concerns can be addressed through design, whereas others can be addressed by providing a clear sequence of steps to eliminate these safety risks. Using the hierarchy of controls (NIOSH, 2024), design features that can be added or eliminated to improve safety should be given priority, followed by any sequence changes. Furthermore, instructions and warnings can be enhanced so that users can clearly understand the potential risks and use appropriate caution.

      For instance, instructions for using the exoskeleton should clearly specify whether any of the pre-use or post-use conditions require two-person operation or assistance from others. This is particularly true if a two-person operation is required to prevent safety risks for users with balance, mobility, or other problems, particularly when donning and doffing exoskeletons. If possible, the need for a two-person operation should be eliminated or reduced through design

features. Otherwise, clear instructions specific to the steps requiring assistance should be provided.

Finally, the design of an exoskeleton should consider factors that are important for in situ use. Factors such as assembly and storage space, training needs, the need for additional assistance to wear and remove the exoskeleton, and the increasing diversity of the worker population that the exoskeletons serve could all be important determinants of future adoption by workplaces. If design features and tasks during pre-use, use, and post-use are amenable to these factors, adoption and implementation can become seamless.

## 5. Conclusions and Future Work

We recommend the following important future directions for research: (1) evaluation of exoskeletons should include safety and accessibility evaluations in addition to usability evaluations; (2) evaluation of exoskeletons with potential as future PPE should be considered; (3) the use of exoskeletons in conjunction with assistive devices should be investigated; and (4) evaluation of exoskeletons with a diverse set of workers, including workers with disabilities, older workers, and workers with temporary mobility impairments should be conducted.

We believe that for exoskeletons to be beneficial, both accessibility and usability evaluations must be conducted to provide design feedback to ensure that all users can benefit from exoskeletons. Furthermore, we urge researchers to conduct safety evaluations of exoskeletons under both use and non-use conditions to address any concerns prior to large-scale implementation in the industry. Future research should also consider the design and process factors that deem a device to be personal protective equipment and understand how these factors are currently considered in exoskeleton design. An important future need is to understand how exoskeletons can be used in conjunction with other assistive devices, such as canes, wheelchairs, or screen readers, so that all users can use exoskeletons. Additionally, assistive devices may need to be evaluated to determine how well they support exoskeleton interactions, potentially leading to the need for new assistive devices. Future studies should also consider a wide range of user populations in user testing studies, including people with disabilities, so that we can understand the design needs and provide scientific evidence for more equitable exoskeleton designs. Currently, there is a dearth of detailed statistics on the types of work and employment performed by persons with disabilities in various industries. The collection and reporting of such data would be beneficial for informing the design of exoskeletons to target specific user groups. We believe that exoskeletons hold plenty of promise for reducing injuries in the workplace and could transform the workplace of today, if it is *designed for all*.